\documentstyle[12pt]{article}
\begin{document}
\rightline{\vbox{\baselineskip12pt\hbox{EFI-95-69}
\hbox{hep-th/9510204}}}
\vskip 1cm
\centerline{{\Large \bf
Integrable Structures in Supersymmetric }}
\vskip .3cm
\centerline{{\Large \bf
Gauge and String Theory}}

\vskip 1.5cm
\centerline{{\Large Emil J. Martinec}}
\vskip .5cm
\centerline{\it Enrico Fermi Institute and Department of Physics}
\centerline{\it University of Chicago,
5640 S. Ellis Ave., Chicago, IL 60637 USA}

\hyphenation{Knizh-hik}
\hyphenation{Zamo-lod-chikov}
\hyphenation{Ber-nard}

\def\pref#1{(\ref{#1})}

\def\ie{{\it i.e.}}
\def\eg{{\it e.g.}}
\def\cf{{\it c.f.}}
\def\etal{{\it et.al.}}
\def\etc{{\it etc.}}

\def\inbar{\,\vrule height1.5ex width.4pt depth0pt}
\def\IC{\relax\hbox{$\inbar\kern-.3em{\rm C}$}}
\def\IR{\relax{\rm I\kern-.18em R}}
\def\IP{\relax{\rm I\kern-.18em P}}
\def\Z{{\bf Z}}
\def\A{{\bf A}}
\def\B{{\bf B}}
\def\Pone{{\bf P^1}}
\def\gg{{\bf g}}
\def\hh{{\bf h}}
\def\ee{{\bf e}}
\def\aa{{\bf v}}
\def\omegat{{\tilde\omega}}
\def\h{{h_\gg}}
\def\hdual{{h_\gg^\vee}}
\def\jac{{\sl Jac}}
\def\One{{1\hskip -3pt {\rm l}}}
\def\sutwo{{$SU(2)$}}
\def\nth{$n^{\rm th}$}

\def\xtilde{{\tilde X}}
\def\pic{{\sl Pic}}
\def\hom{{\sl Hom}}
\def\prym{{\sl Prym}}

\def\beq{\begin{equation}}
\def\eeq{\end{equation}}

\def\sst{\scriptscriptstyle}
\def\tst#1{{\textstyle #1}}
\def\frac#1#2{{#1\over#2}}
\def\coeff#1#2{{\textstyle{#1\over #2}}}
\def\half{\frac12}
\def\hf{{\textstyle\half}}
\def\ket#1{|#1\rangle}
\def\bra#1{\langle#1|}
\def\vev#1{\langle#1\rangle}
\def\d{\partial}

\def\np{{\it Nucl. Phys. }}
\def\pl{{\it Phys. Lett. }}
\def\pr{{\it Phys. Rev. }}
\def\ap{{\it Ann. Phys., NY }}
\def\prl{{\it Phys. Rev. Lett. }}
\def\mpl{{\it Mod. Phys. Lett. }}
\def\cmp{{\it Comm. Math. Phys. }}
\def\grg{{\it Gen. Rel. and Grav. }}
\def\cqg{{\it Class. Quant. Grav. }}
\def\ijmp{{\it Int. J. Mod. Phys. }}
\def\jmp{{\it J. Math. Phys. }}
\def\nextline{\hfil\break}
\catcode`\@=11
\def\slash#1{\mathord{\mathpalette\c@ncel{#1}}}
\overfullrule=0pt
\def\AA{{\cal A}}
\def\BB{{\cal B}}
\def\CC{{\cal C}}
\def\DD{{\cal D}}
\def\EE{{\cal E}}
\def\FF{{\cal F}}
\def\GG{{\cal G}}
\def\HH{{\cal H}}
\def\II{{\cal I}}
\def\JJ{{\cal J}}
\def\KK{{\cal K}}
\def\LL{{\cal L}}
\def\MM{{\cal M}}
\def\NN{{\cal N}}
\def\OO{{\cal O}}
\def\PP{{\cal P}}
\def\QQ{{\cal Q}}
\def\RR{{\cal R}}
\def\SS{{\cal S}}
\def\TT{{\cal T}}
\def\UU{{\cal U}}
\def\VV{{\cal V}}
\def\WW{{\cal W}}
\def\XX{{\cal X}}
\def\YY{{\cal Y}}
\def\ZZ{{\cal Z}}
\def\lam{\lambda}
\def\eps{\epsilon}
\def\vareps{\varepsilon}
\def\underrel#1\over#2{\mathrel{\mathop{\kern\z@#1}\limits_{#2}}}
\def\lapprox{{\underrel{\scriptstyle<}\over\sim}}
\def\lessapprox{{\buildrel{<}\over{\scriptstyle\sim}}}
\catcode`\@=12

\begin{abstract}
The effective action of N=2 Yang-Mills theory with adjoint matter
is shown to be governed by an integrable spin model with spectral
parameter on an elliptic curve.  We sketch a route to deriving
this effective dynamics from the underlying Yang-Mills theory.
Natural generalizations of this structure to all N=2 models, and
to string theory, are suggested.
\end{abstract}

\vskip .5cm

\section{Introduction}

Recent results in N=2 supersymmetric
gauge theory \cite{russians,MW,NT,DW}
suggest an intimate connection between the low-energy effective theory
and integrable systems.  In hindsight, this connection is
natural given the data that specifies the effective
theory: The moduli space of vacua is parametrized
by fundamental Casimirs of the Lie algebra; and the BPS mass formula
$M(\vec m,\vec n)=|\vec m\cdot \vec a + \vec n\cdot \vec a_D|$
respects a natural $Sp({\rm rank}(G),\Z)$ symmetry -- where
the normalizations $a$, $a_D$ in this formula may be
represented by periods of a special differential
$\lam_{SW}$ on a Riemann surface
$\Sigma$\cite{SW}, whose moduli space is that of the
gauge theory.  On the other hand, precisely the
same data characterizes the simplest
integrable systems with periodic motions;
a set of integrals of motion in one-to-one correspondence
with the Casimirs of a Lie algebra, and evolution
that linearizes via the Liouville map on the Jacobian of a
Riemann surface $\Sigma$ (the spectral curve), whose moduli are the
integrals of motion.  Moreover, the action coordinates are the
periods of a special differential $pdq$ around particular cycles
(and of course the periods respect a natural symplectic structure).
This connection has been made in pure N=2 Yang-Mills theory
for all simple gauge groups in \cite{MW}.
The relevant integrable system is the twisted affine Toda lattice.

This reasoning strongly suggests that one should be able to find
an integrable system for any N=2 model.  When matter is
included, the data for the
integrable system must be enlarged to include the masses of
matter multiplets, which enter into the algebraic equation for
the spectral curve in a particular way\cite{SW2,HO,APS,AS,hanany,DW}.
In a recent work, Donagi and Witten\cite{DW} showed that,
in an ultraviolet finite theory -- namely softly
broken N=4 Yang-Mills -- there is an integrable system with
spectral parameter
associated to a torus whose modulus is the microscopic
gauge coupling $\tau=\frac{\theta}{2\pi}+\frac{4\pi i}{e^2}$.
Here we will connect these results
with those of the author and Warner\cite{MW}.   Specifically,
the integrable system of \cite{DW} is in fact the elliptic
spin model of Calogero-Moser-Sutherland-Olshanetsky-Perelomov
(see for instance \cite{OP}).  It is known\cite{inoz-toda}
that this system degenerates to the periodic Toda lattice
in a certain limit.  There is a version of this integrable system
for any Lie group\cite{OP}; in fact there exists a multiparameter
deformation of the potential which is known to preserve
integrability\cite{inoz-lax}, and which can degenerate to a variety
of trigonometric integrable systems\cite{inoz-toda}
as one degenerates the spectral parameter torus.
We will describe this elliptic spin model in section 2.
In section 3 we will sketch an idea of how to connect the
integrable potential to the underlying Yang-Mills theory.
This is followed by a conjecture about how to construct
an integrable system for any N=2 theory using superalgebras.
In section 4 a natural role of the quantized integrable system
is explored.  Finally, using recent results
of Harvey and Moore\cite{HM}, we suggest in section 5 that
an extension of these ideas to hyperbolic algebras should yield an
integrable system governing nonperturbative N=2 string theory.

\section{Elliptic spin models}

A central point of Donagi and Witten's work\cite{DW} is that finite
N=2 theories should have a spectral parameter $z$ living on an elliptic
curve.  One then imagines the pure N=2 theory arising from
the degeneration of the elliptic curve to a twice-punctured sphere,
perhaps while simultaneously shifting some variables.
In fact, there is an extremely natural integrable system
with this property: The elliptic Calogero-Moser model
(the system of type IV in the classification of Olshanetsky
and Perelomov\cite{OP}).  It has as simplest (quadratic)
integral of motion
\beq
H_2=\hf p^2 + {\hf}\sum_{\alpha} g^2 \wp(\alpha\cdot q|\tau)
\label{wp-ham}
\eeq
\noindent
where $g$ is a coupling constant (we will soon see it is the
mass of the adjoint hypermultiplet in \cite{DW}), $\alpha$ are
the roots of a Lie algebra $\bf g$, and
$\wp$ is the Weierstrass function.  There are rank($\gg$) canonical
pairs of dynamical variables $p_i$, $q_i$.  The $q_i$ are
the vevs of the adjoint Higgs field of the gauge multiplet.
Let the periods of the torus be $\omega_1=2\pi i$ and $\omega_2$,
with $\tau=-\omega_2/\omega_1$.
The reduction to affine Toda is accomplished by the limit\cite{inoz-toda}
\beq
\omega_2\rightarrow \infty\quad , \qquad q_j=x_j + c_j\omega_2 \quad ,
\qquad g^2=e^{b\omega_2}\ .
\eeq
(at least for the two cases that have been studied in detail in the
literature: groups $A_n$ and $D_n$, where $c_j=(j-1)/\h$ and
$b=1/\h$; here $\h$ is the Coxeter number, which is the same as
$\hdual$ for these groups).
The dynamics collapses from interactions on
the entire root lattice down to
exponential interactions for
the simple roots (including the affine root) due to the
asymptotics (and periodicity) of the Weierstrass function.
The limit gives affine Toda with
dynamical pair $p_i$, $x_i$.  The other groups have not yet been checked
to see that they reduce properly to affine Toda dynamics on the dual group
needed for pure N=2 gauge theory; however, there seem to be enough
parameters in the general potentials of \cite{inoz-toda} to
accomodate the Toda theories based on twisted Kac-Moody algebras.
One might also generate the twisted affine Toda models using
the orbifold constructions of \cite{OT}.
We do not regard these approaches as particularly natural.
Rather, if the promising ideas of section 4 bear fruit, the dual group
will arise automatically and in an entirely different way.

Olshanetsky and Perelomov\cite{OP}
give a Lax pair formulation with Lax operator
\begin{eqnarray}
L=\sum_{i} p_i \hh_i + g \sum_{\alpha} \phi(\alpha\cdot q,z) \ee_\alpha
\nonumber\\
\phi(x,z)=\frac{\sigma(x-z)}{\sigma(x)\sigma(z)} exp[\zeta(z) x]
\label{lax}
\end{eqnarray}
where $\sigma$ and $\zeta$ are the corresponding Weierstrass functions
and $z$ is the spectral parameter.
Note that the Lax operator
has very much the structure required by Donagi and Witten:
holomorphic except at $z=0$ where it has a simple pole; the
residue is the operator $\sum_\alpha \ee_\alpha$ which for $A_n$
has eigenvalues $(1,1,...,1,-(n-1))$
when written as a matrix in the fundamental
representation\footnote{I thank E. Witten
for pointing out the relation between the residue of \pref{lax}
and the section of the Higgs bundle in \cite{DW}.}.
The overall coefficient of this residue was argued in
\cite{DW} to be the mass $\mu$ of the adjoint hypermultiplet;
hence we identify $g$ as that mass.
This system fits rather nicely into the Hitchin framework
as well, \cf\ \cite{nekrasov,olsh}.
In fact, Nekrasov\cite{nekrasov} and Olshanetsky\cite{olsh} have
shown that the classical Hitchin system on an elliptic curve with
bundle monodromy $diag(\exp[2\pi i  q\cdot\hh])$
generates a (Hitchin system, not Yang-Mills)
Higgs vev which is essentially \pref{lax},
when one chooses an appropriate coadjoint orbit at the
pole (related to a certain symmetric space\cite{OP,ABT,KBBT}).

\section{A route to integrability?}

The finite N=2 models are special in that one still has control
over the microscopic gauge coupling in the effective theory;
the action is not renormalized.  Thus one might imagine
deriving the elliptic spin model by considering the dynamics
of the low-energy degrees of freedom.  In particular,
much of the structure of the low-energy theory seems
to be governed by the fermion zero-modes.  For instance,
in asymptotically free theories they determine
the number of vacuum states, the discrete breaking of $R$-symmetry,
and so on.  In the analogous two-dimensional N=2 models,
an important role is played by the fermion zero-mode
Hilbert space as a function of the parameters\cite{CGP,CV}.
Thus it is intriguing that terms can be generated
in the effective potential for the fermions which have very
much the structure of the potential in \pref{wp-ham}.
Consider integrating out the massive vector multiplets of the
root generators of $\gg$ in
the Yang-Mills action.  Their masses (\eg\ think of the scalar component)
come from couplings like $tr\{[\phi^\dagger,\phi]^2\}$; the mass of
$\phi_{\alpha}$ is roughly $(\alpha\cdot v)^2$, where $v=\vev\phi$
(it is perhaps better to take $v$ real for the moment).
Now let us consider the effect on the low-energy fermion
modes of integrating out the massive vector multiplets.
This is induced by the Yukawa coupling
$\phi_\alpha \bar\psi t_{-\alpha}\psi$.
If we call $\bar\psi t_\alpha \psi = S_\alpha$, integrating
out $\phi_\alpha$ induces a potential
\beq
\sum_{\alpha} \frac{S_\alpha S_{-\alpha}}{(\alpha\cdot v)^2}
\label{handwave}
\eeq
This should be accurate for weak Yang-Mills coupling and small
Higgs vev $v$.  However we know how to extend this to a more
precise answer.  Recall the work of Harvey \etal\cite{HMS}
on the dimensional reduction of super-Yang-Mills: If we consider
the $N=4$ theory to come from higher dimensions, compactified
on a torus, then $\alpha\cdot v$ should be periodically identified
due to gauge transformations in the higher-dimensional theory.
Then the potential \pref{handwave} is that of the spin generalization
of the trigonometric
Calogero-Moser-Sutherland system (\cf\ \cite{nekrasov,KBBT})!
That is, if the spin state is one with symmetry
$S_\alpha=const.$ independent of $\alpha$,
then one obtains the ordinary Calogero-Moser-Sutherland
potential $(1/\sin^{2})$.  Finally, if we assume that the spectral
parameter of our system lives on a torus, the above potential
should be further generalized to a doubly-periodic function.
Perhaps this periodic structure will come from a better understanding of
the field space of the adjoint Higgs away from the origin.
In any event, the natural setting seems to include a set of fermions
living on the spectral parameter torus.
That is, one imagines a fiber bundle with fiber the Hilbert space
of fermion zero modes, and base space the total space of
the universal elliptic curve
(parametrized by the spectral variable $z$)
over the moduli space of Higgs vevs $v$ and gauge couplings $\tau$;
one would then try to perform an analysis along the lines of
\cite{CGP,CV}\footnote{This line of reasoning was proposed already
in \cite{MW}.}.

We now see a possible route to getting all finite N=2 theories:
Consider N=2 Yang-Mills coupled to an appropriate combination of
hypermultiplets in representations $R_i$ of $G$, yielding a finite theory.
Integrating out the massive vector multiplets will yield
a potential exactly as above, except that $S_\alpha$ has
a sum of contributions from the generator $S_\alpha(R_i)$ of each
hypermultiplet.  One thing that needs to be explained
is how vector multiplet fermions and hypermultiplet
fermions enter with opposite signs in situations like the
perturbative beta function $\beta_\tau\propto [2C_2(G)-\sum_i T_2(R_i)]$.
One might then understand the finiteness of the spacetime N=2
theory as the vanishing of a Chern class or holomorphic anomaly
of fermions living
on the spectral parameter curve; for instance, the fermion currents
$S_\alpha$ have a current algebra anomaly $T_2(R_i)$ (equals $C_2(G)$
for the adjoint representation) if they
are thought of as two-dimensional fermions on the spectral
parameter curve.

A somewhat different approach is suggested by the string theory
work of Harvey and Moore\cite{HM}.  There, a superalgebra
was defined on BPS states, with the even generators corresponding
to vector multiplets, and the odd generators
corresponding to hypermultiplets.
This indicates that one might consider an integrable system
based on a Lie superalgebra for the N=2 theories with matter.
This is particularly natural if one follows the logic of
Seiberg\cite{nati-NR} and regards the hypermultiplet masses as
arising from the expectation value of a field.  Indeed,
one may gauge the flavor group\footnote{I thank
N. Warner for discussions about gauging flavor symmetry
in N=2 theories.}
but look in the weak-coupling
limit ($e_{\sst flavor}\rightarrow 0$ for finite theories,
$\Lambda_{\sst flavor}\rightarrow 0$ for asymptotically free ones);
then the $N_f$ hypermultiplet masses are the flavor group Higgs vevs,
but the flavor dynamics is decoupled (the corresponding spectral
curve is completely degenerate, so only poles will arise in
the integral that normalizes the part of the BPS mass formula
coming from hypermultiplet masses).
Of course, one need not specialize to this limit; in general
there will be an interesting interplay between
the spectral covers for the two groups.
In the end, one is looking for superalgebras whose bosonic part
is $G_c\times G_f$ and whose odd part transforms in the proper
representations of each group.
Thus for $SU(N_c)$ with $N_f$ fundamental flavors, one should
consider the supergroup $SU(N_c|N_f)$; for $SO(N_c)$ gauge theory,
$OSp(N_c|N_f)$; and for $Sp(N_c)$, gauge theory $OSp(2N_f|N_c)$.
Higgs branches of the moduli space of vacua of the N=2 gauge
theory should be realized when one puts nontrivial holonomy
(of the Hitchin Higgs bundle over the spectral parameter torus)
in the odd directions of the group, corresponding to giving
a vacuum expectation value to a hypermultiplet.
Similarly, the adjoint hypermultiplet of the softly broken
N=4 theory ought to be a fermionic component of the
Hitchin system of the previous section.  A logical candidate
is the supersymmetric version of Hitchin's construction,
with the bosonic fields related to the adjoint vector multiplet
and their superpartners to the adjoint hypermultiplet.
It is not immediately clear what this ansatz has to do with
the above analysis of the low-energy fermion dynamics.

Thus the use of superalgebras opens the door to a unified
description of the Higgs and abelian Coulomb phases.
It is hoped that a combination of the ideas presented here
and those of \cite{SW,SW2,DW} will enable a unified
description of all phases of supersymmetric gauge theory.
Control over the microscopic couplings provided by finiteness
may also lead to a proof of N=1 dualities\cite{seib_Neq1}.

\section{Quantization of the spin model}

One of the key observations of \cite{russians,MW,NT} is that
the effective prepotential of pure N=2 Yang-Mills gauge theory
is the free energy of the {\it Whitham averaged} hierarchy
of the underlying integrable system.  The averaging procedure
is an adiabatic deformation dynamics on the moduli space of the
original integrable system (in this case twisted affine Toda).
An alternate perspective on this procedure is that it is the
first step in the WKB quantization of the integrable
system\cite{DoMa}.  This suggests that, to some degree, we should
not be considering the classical integrable system but
rather its quantum counterpart.  The quantization of the
spin generalization of the Hamiltonian \pref{wp-ham}
has an extremely interesting interpretation:
It is essentially the RHS of the Knizhnik-Zamolodchikov-Bernard
equation for the conformal blocks of the WZW theory on the
once-punctured torus.  More precisely, the KZB equation in
this case is\cite{FW}
\beq
4\pi i(k+\hdual)\frac{\d}{\d\tau}\omegat(z,\tau,\vec q) =
\frac{\d^2}{\d \vec q^{\; 2}}\omegat - \eta_1(\tau) C_2(V)\omegat
-\sum_{\alpha\in\Delta}\wp(\vec\alpha\cdot \vec q)
\ee_\alpha\ee_{-\alpha}\omegat
\eeq
where $\omegat(z,\tau,\vec q)=\Pi(\tau,\vec q)\omega(z,\tau,\vec q)$
with $\Pi$ the Weyl-Kac denominator, $\omega$ a conformal
block of the once-punctured torus with representation $V$ at the
puncture; $\eta_1(\tau)$ is the first nonpole term in the expansion
of $\d_z\log[\vartheta_1(z|\tau)]$ near $z=0$.
At the critical level $k=-\hdual$, the KZB equation coincides with
the action of the quadratic quantum Hamiltonian of the Hitchin system
on the quantum Hilbert space\cite{nekrasov,olsh}.
In this case, the Hamiltonian (and other integrals of motion)
falls into the center $\CC_{-\hdual}$ of the universal enveloping
algebra $U_k(L\gg)$ at $k=-\hdual$ \cite{BD,FeFr,ER}.
The analogue of this center when $k\ne\hdual$ is the
$W$-algebra $W_k(\gg)$ \cite{FeFr}.
There is a remarkable duality
$W_k(\gg)\approx W_{k^\vee}(\gg^\vee)$ \cite{FeFr},
related to the geometric Langlands program \cite{BD}.
Here $r^\vee(k+\hdual)=(k^\vee+h_{\gg^\vee}^\vee)^{-1}$, with $r^\vee$
the maximal number of edges connecting two vertices of the
Dynkin diagram of $\gg$ (\ie\ the order of the diagram automorphism
in the orbifold construction of twisted Kac-Moody algebras).
Thus it would seem that the WZW model near the critical
level (related to the quantized Hitchin system), is related to
the WKB limit $k\rightarrow \infty$ of the dual Kac-Moody;
which may explain the appearance of WKB averaged, {\it twisted}
affine Toda in the pure gauge N=2 theory\cite{MW}
if one can prove that the partition function of the
finite theory satisfies the KZB equation in the critical level limit.
This duality is also natural when one considers that the
ultraviolet limit of the Yang-Mills theory is N=4 with gauge group $G$
(and integrable system elliptic Calogero-Moser on $G$),
whereas the infrared theory is an N=2 gauge theory of monopoles
with gauge group $G^\vee$ (and integrable system
affine Toda on $(LG)^\vee$).  The transition between the two
descriptions should be made natural by the renormalization
group flow corresponding to the range of theories with
adjoint masses between $\mu=0$ and $\mu=\infty$.
It would be amusing if, as appears to be the case, Montonen-Olive
and Langlands duality were one and the same.

\section{Extension to string theory}

Recent work of Harvey and Moore\cite{HM} indicates that the
above ideas have an elegant and extremely natural generalization
to string theory.  These authors found that the structure of
threshold corrections in the N=2 Heterotic string
(\eg\ compactified on $K3\times T^2$) were organized by
the structure of a generalized or hyperbolic Kac-Moody (super)algebra.
At first sight one might doubt that the above setting could
be carried over to string theory, because the beta function
of the theory does not vanish.  Nevertheless, the theory is finite
(because it's string theory), and one might regard the breaking
of supersymmetries by the compactification manifold to be a
soft breaking (it is for instance restored in the large
radius limit)\footnote{Also, even in nonsupersymmetric solutions
the spectrum looks asymptotically supersymmetric\cite{KS}.}.
The complex coupling constant of the theory is the expectation
value $S$ of the dilaton multiplet.  Thus we should look for
an integrable system on a spectral parameter torus with
modulus $\tau=S$.  The gauge group has been identified by
Harvey and Moore\cite{HM}: a hyperbolic Kac-Moody (super)algebra
acting on the BPS states.  The proof of integrability of models
like \pref{wp-ham},\pref{lax} relies on little more
than that $\ee_\alpha$, $\hh_i$ are the generators of a Lie algebra,
and so should hold for the hyperbolic case.  The dynamical coordinates
of the integrable system will be the Higgs vevs $\vec v$ in the CSA
(expectation values of Wilson lines in \cite{HM}), together with
the expectation values $T$, $U$ of the Narain moduli corresponding
to the additional Cartan generators of the hyperbolic algebra.
Thus we conjecture that \pref{lax} will be the Lax operator
of the integrable system governing nonperturbative
N=2 Heterotic string theory, with
$\vec q=(T,U,\vec v)$, $\tau=S$, and $\gg$ the string
gauge (super)algebra.  Using the odd generators of the
superalgebra, again
one might reduce the rank of the low-energy gauge group
by passing to a Higgs branch along the lines suggested in section 3.

\vskip 1cm
\noindent {\bf Acknowledgements:} I am grateful to
J. Harvey for discussions and for communicating the results of \cite{HM}
prior to publication; to E. Witten for discussions; and especially
to N.P. Warner for discussions and ongoing collaboration which led
to this work.
This work is supported in part by funds provided by the DOE under
grant DE-FG02-90ER-40560.

\end{document}